# RheoSpeckle: a new tool to investigate local flow and microscopic dynamics of soft matter under shear

N. Ali[1], DCD. Roux[1], L. Cipelletti[2], and F.Caton[1]

[1] Univ. Grenoble Alpes, LRP, F-38000 Grenoble, France
CNRS, LRP, F-38000 Grenoble, France
[2] Laboratoire Charles Coulomb (L2C), UMR 5221 CNRS-Université de Montpellier, F-34095 Montpellier, France

E-mail: nabil.ali@univ-grenoble-alpes.fr, denis.roux@univ-grenoble-alpes.fr

**Abstract**
To investigate the interplay between microscopic dynamics and macroscopic rheology in soft matter, we couple a stress-controlled-rheometer equipped with a Couette cell to a light scattering setup in the imaging geometry, which allows us to measure both the deformation field and the microscopic dynamics. To validate our setup, we test two model systems. For an elastic solid sample, we recover the expected deformation field within 1 µm. For a pure viscous fluid seeded with tracer particles, we measure the velocity profile and the dynamics of the tracers, both during shear and at rest. The velocity profile is acquired over a gap of 5 mm with a temporal and spatial resolution of 1s and 100 µm, respectively. At rest, the tracer dynamics have the expected diffusive behavior. Under shear, the microscopic dynamics corrected for the average drift due to solid rotation scale with the local shear rate, demonstrating that our setup captures correctly the relative motion of the tracers due to the affine deformation.

## 1 Introduction

Many complex fluids and viscoelastic systems of industrial and biological relevance including gels, concentrated colloidal suspensions or emulsions, surfactant phases, composite materials, biogels, and biological fluids such as mucus are heterogeneous on length scales of the order of 1 µm – 100 µm. The impact of this heterogeneity on the macroscopic rheology is currently of great interest [1] [2] [3], both at a fundamental level and for industrial applications. Indeed, a complex fluid may be characterized by the existence of a relevant mesoscopic length scale in between the size of individual molecules and the size of the sample [4]. The existence of such microstructure can lead to a very rich mechanical behavior, as exemplified by the non-monotonic relation between an applied shear stress and the resulting shear deformation, e.g. in wormlike micelle solutions [5], lyotropic lamellar phases [6] or, quite generally, in yield stress fluids [1] [7].

To fully understand these behaviors, these materials should be probed not only at the macroscopic scale, as in conventional rheology, but also at the scale of their heterogeneity. During the last years, several non-destructive techniques have been introduced to investigate heterogeneous flows at small scale, such as: Particle Imaging Velocimetry PIV [8], laser doppler velocimetry LDV [9] and speckle-based ultrasonic velocimetry methods [10] [11]. All these powerful techniques use micron-sized tracer beads that may perturb the sample structure. Nuclear Magnetic Resonance (NMR) [12] [13] is an alternative appealing, non-invasive method, but it requires the use of powerful magnets and thus remains costly and difficult to implement.

Here, we probe the deformation field and the microscopic dynamics of a mechanically driven material using non-conventional, space-resolved dynamic light scattering. We work in the single scattering regime; accordingly, only minute amounts (volume concentration ~ $10^{-5}$) of tracer particles on the nanometric scale have to be added to the sample. In fact, adding tracers is not even strictly required, provided that the sample displays sufficient refractive-index fluctuations, as recently shown in a microscopy-based apparatus [14]. The setup couples a speckle imaging apparatus, described in detail below, to a MCR301 stress-controlled rheometer (Anton Paar) equipped with a transparent Couette cell. In brief, the sample is illuminated by a laser beam that crosses the Couette geometry perpendicularly to the rotation axis. The laser light is scattered by the tracer particles, thereby creating an interference pattern termed speckle image [15]. The optical layout is inspired by the imaging geometry proposed by Duri et al [16]: each portion of the speckle image corresponds to a well-defined zone of the illuminated sample. Accordingly, the motion of the sample, e.g. under shear, is reflected by a displacement of the speckle field that can be precisely measured. In addition to the drift motion, the speckles exhibit intensity fluctuations that are associated to the relative motion of the tracers, thereby providing valuable information on the microscopic dynamics of the sample. These intensity fluctuations are analyzed using the multispeckle variant of dynamic light scattering [17] [18] [19]. Dynamic light scattering probes the dynamics on a length scale of the order of the inverse scattering vector $q = 4\pi n \lambda^{-1} \sin\theta/2$, where $n$ is the solvent refractive index, $\lambda$ is the in-vacuo laser wavelength and $\theta$ is the scattering angle.

The paper is organized as follows. In Section 2, we first introduce briefly the tested systems. We then describe the setup, we an emphasis on the choice of the optimum speckle size. Section 3 introduces the various algorithms for measuring the deformation field and the microscopic dynamics corrected for the contribution of the average drift. A test of the algorithm on synthetically generated speckle patterns is also shown here. Section 4 deals with a series of tests of the new apparatus, on a purely elastic solid and on a Newtonian fluid. For the former, the measured deformation field in the elastic regime is successfully compared to the expected one. For the latter, we show that the velocity field under shear and the microscopic dynamics of the tracer particles, both at rest and under shear, can be precisely measured by the new setup. Section 5 recapitulates the main results.

## 2   Materials and Experimental setup

### 2.1   Materials

#### 2.1.1   Elastic solid

Cross-linked polydimethylsiloxane (PDMS, Sylgard 184) is prepared by mixing a cross linker containing spherical silica particles of 300 nm diameter (manufactured by EPREI) with the elastomer at a ratio of 10:1. The sample is then degassed in a vacuum chamber. Finally, the sample is cured in an oven at 60°C for a few hours. The cross-linked PDMS behaves mechanically as an elastic solid (See Rheological curves in the SI).

#### 2.1.2   Viscous fluid

The Newtonian sample is a dispersion of a polystyrene microspheres (diameter $D_h$=180 nm, measured using a commercial Malvern dynamic light scattering (DLS) instrument) in a 58% weight fraction aqueous solution of EmkaroxHV45, a high viscosity polyalkylene glycol. The sample is purely Newtonian (Rheological curves in the SI) with viscosity $\eta = 11.4 Pa.s$ at 24°C. The particle volume fraction is fixed at C=5×10$^{-5}$ in order to be in the single scattering condition.

### 2.2   Apparatus

A sketch of the RheoSpeckle experimental setup is shown in Figure 1. A MCR301 stress controlled-rheometer equipped with a transparent Couette plexiglass cell (internal radius $R_i=15$ mm, external radius $R_e=20$ mm) is coupled with a speckle-imaging system. A linearly polarized (along the $y$ axis), single-mode laser (MSLIII, CNI, China, $\lambda=532nm$), is used to illuminate the sample in the plane defined by the velocity and velocity gradient directions. A laser sheet is produced using two lenses. The first ($f_{l1}'=200mm$) focuses the laser in the observation plane. The second one (cylindrical lens, $f_{l2}'=-25.4mm$) is placed at 65 mm from the first lens, and produces a 5mm wide and 500 μm thick laser sheet. The coherent light is scattered by the tracer particles (or by the fluctuations of refractive index inherent to the sample). The light scattered at 90° is reflected by a non-polarizing mirror before being collected by a lens with focal length $f_{l3}'=150mm$. A diaphragm placed in the focal plane of the lens allows adjusting the speckle size as explained hereafter. The role of this lens is to form an image of the scattering volume onto the sensor of a CCD camera (acA640-100gm, Basler, Germany). The magnification value $G=0.5$ is chosen so as to image the full gap onto the detector area (2.8×3.7 mm²). The collected light is scattered at a scattering vector $q \sim 22$ μm, thus probing the microscopic dynamics of the sample on a length scale $q^{-1} \sim 50$ nm [20].

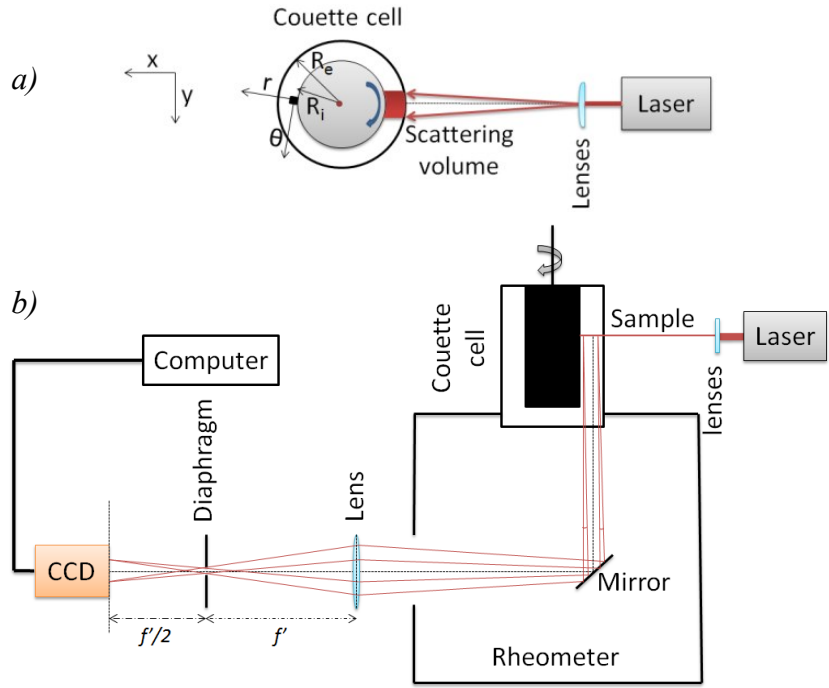

Figure 1. (Color online). Sketch of the experimental set-up (RheoSpeckle). a) Top view, b) Side view.

## 2.3 Choice of the speckle size

The diaphragm shown in Figure 1(b) allows the average size of the speckles $S$, to be adjusted. In this configuration, the size of the speckles is inversely proportional to the diaphragm aperture $D$ and depends on the wavelength $\lambda$, the magnification value $G$ and the distance between the diaphragm and the image plane $d$ via [21]

$$S = 2.4 \frac{\lambda G d}{D} \quad (1)$$

If the speckle size is significantly smaller than the pixel size, the speckle contrast decreases as the square root of the number of coherence regions inside the measurement area [22]. A low contrast makes the analysis of the speckle images difficult. On the other hand, however, the number of collected speckles should be maximized, in order to achieve the best averaging. This is particularly important since the image will be divided in smaller portions to be independently processed in order to gain insight on the local properties of the sample. It is therefore important to adjust the speckle size in order to strike the best compromise between a good contrast and a large number of collected speckles.

To investigate this issue, we measure the speckle size of frozen speckle patterns produced by the seeded PDMS. A series of speckles images scattered by a volume of $5 \times 5 \times 0.5$ mm$^3$ are taken for a series of diaphragm apertures, ranging from 2 to 6 mm. We quantify the speckle size by calculating the spatial autocorrelation function $corr[I,I](k,l)$ of the images taken with different diaphragm apertures, where

$$corr[I,I](k,l) = \frac{\text{cov}[I,I](k,l)}{\text{var}[I]} \qquad (2)$$

$$\text{cov}[I,I](k,l) = \frac{1}{N}\sum_{r,c} I_{r,c} I_{r+k,c+l} - \frac{1}{N^2}\sum_{r,c} I_{r,c} \sum_{r,c} I_{r+k,c+l} \qquad (3)$$

$$\text{var}[I] = \frac{1}{N}\sum_{r,c} I^2_{r,c} - \left(\frac{1}{N}\sum_{r,c} I_{r,c}\right)^2 \qquad (4)$$

Here, $I_{r,c}$ is the intensity at a time $t$ of the pixel at row $r$ and column $c$ and $k$ and $l$ are spatial shifts expressed in number of rows and columns, respectively. We fit the autocorrelation function with a Gaussian curve characterized by the full width $S$ at half height. The contrast of the speckle pattern corresponds to the value of $corr[I,I](x=0, y=0)$ and the noise level is the standard deviation of the base line of $corr[I,I]$ (see inset of Figure 2(a)).

Figure 2(a), shows the speckle size $S$ plotted as a function of diaphragm aperture $D$, showing good agreement between theory (Eq.(1)) and experiments. Figure 2(b) shows the contrast and noise levels as a function of the speckle size. The contrast increases with speckle size and a maximum is reached for a speckle size of 4.5 pixels. However, the noise level is smallest when the speckle size is between 2 and 3 pixels. Hence, we choose as the optimum value of the speckle size 3 pixels, which gives us a good compromise between the contrast (>0.5) and the noise level (<10$^{-2}$). This value of $S$ is consistent with the one found by Viasnoff et al [23] for a multispeckle apparatus for Diffusing Wave Spectroscopy (highly turbid samples) in the far-field transmission geometry.

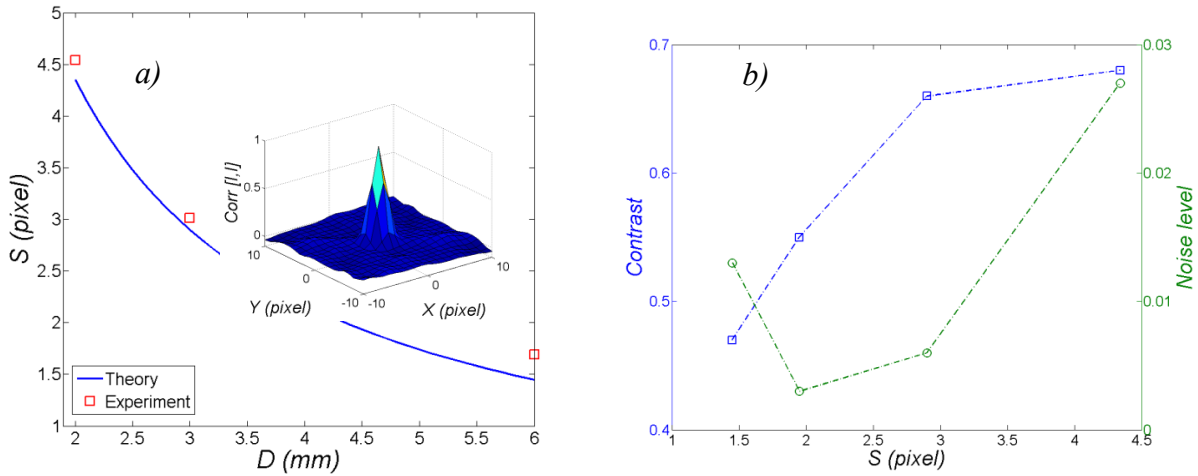

Figure 2. (Color online). a) Measured speckle size as a function of the diaphragm diameter $D$ (squares) and theoretical prediction (line, Theoretical curve from Eq (1). The speckle size is defined as the full width at half

height of a Gaussian fit to the intensity spatial autocorrelation function. Inset: 3D Spatial intensity autocorrelation function for $D = 3mm$ as a function of the spatial shift in number of pixels along $x$ and $y$. b) Contrast (left axis) and noise level (right axis) as a function of the speckle size.

## 3  Space-resolved dynamic light scattering

To probe the microscopic dynamics of materials, we use the space- and time-resolved correlation method of [24] and [15]. The space and time-resolved intensity correlation is defined as follow:

$$c_I(t,\tau) = \frac{\langle I_p(t) I_p(t+\tau) \rangle_r}{\langle I_p(t) \rangle_r \langle I_p(t+\tau) \rangle_r} - 1 \quad (5)$$

In the above equation $I_p$ is the scattered intensity of the $p$-th pixel of a region of interest (ROI) centered around the position $r$, and the averages are taken over all pixels of the ROI. A ROI corresponds to a small volume in the Couette gap, typically of size $0.2 \times 0.2 \times 0.5$ mm$^3$. The average dynamics is quantified by calculating the usual DLS intensity correlation function, expressed in terms of $c_I(t,\tau)$ by:

$$g_2(\tau) - 1 = \langle c_I(t,\tau) \rangle_t \quad (6)$$

One must be cautious in applying Eqs (5) and (6) to a sheared sample. When a sample is sheared in the gap of the Couette geometry, it is submitted to a deformation field which is the sum of a solid rotation and of a stretching. Locally, the solid rotation results in a drift of the sample. As a consequence, the speckle pattern corresponding to a ROI will be progressively shifted. Additionally, as the ROI is stretched, the scatterers are displaced relatively to each other, which results in intensity fluctuations on top of the speckle drift motion. This relative motion cannot be directly obtained by applying Equations 5 and 6, because these correlation functions are sensitive also to the speckle drift, not only to the intensity fluctuations resulting from the relative motion of the scatterers. To measure both the local drift and the microscopic dynamics, we use the method introduced in Ref. [25]. First, the drift is reconstructed from the position of the peak of the spatial crosscorrelation between two images taken at time $t$ and $t+\tau$, as in conventional particle imaging velocimetry (PIV) [26] [27]. Then, to correct for the contribution of drift, we calculate for each ROI $c_I(t,\tau)$ and $g_2(\tau) - 1$ in a reference frame that is co-moving at the same drift velocity as that of the ROI. As shown in [25], such drift-corrected correlation function may be conveniently calculated by applying the following formula:

$$g_2(\tau) - 1 = \frac{\sum_{k,l=-M/2+1}^{M/2} h(\delta y - k) h(\delta x - l) \mathrm{cov}[J,I](k+i_y, l+j_x)}{\overline{J}\,\overline{I}} \quad (7)$$

Here, $I$ is the matrix of intensity values at time $t$, $J$ is the corresponding intensity matrix at time $t+\tau$, $k$ and $l$ are spatial shifts expressed in number of rows and columns, $\overline{I} = N^{-1} \sum_{r,c} I_{r,c}$ and similarly for $\overline{J}$, $h$ is a truncated windowed sinc function with an even number $M$ of elements for evaluating the kernel $h$, $\Delta x = j_x + \delta x$ and $\Delta y = i_x + \delta y$ are the displacement between the two images $I$ and $J$ in the $x$ and $y$ direction, respectively, with $j_x = floor(\Delta x)$ and $i_y = floor(\Delta y)$, where $floor(x)$ designates the smallest integer $\leq x$.

To test the efficiency of the solid drift correction method and to evaluate the best choice of $M$, we numerically apply successive homogeneous displacements(i.e. solid drifts) of 1 pixel into each direction to a real speckle image generated by the cross-linked PDMS at rest.

Figure 3 shows the resulting intensity correlation function as a function of the imposed displacement, using Eq (6) (raw data) or Eq (7) (drift-corrected data). The correlation function computed from the uncorrected data decreases rapidly to zero while the corrected one remains constant for all imposed displacements. These data demonstrate that the drift correction eliminates the contribution due to the solid body motion, thus allowing the microscopic relative motion to be correctly measured.

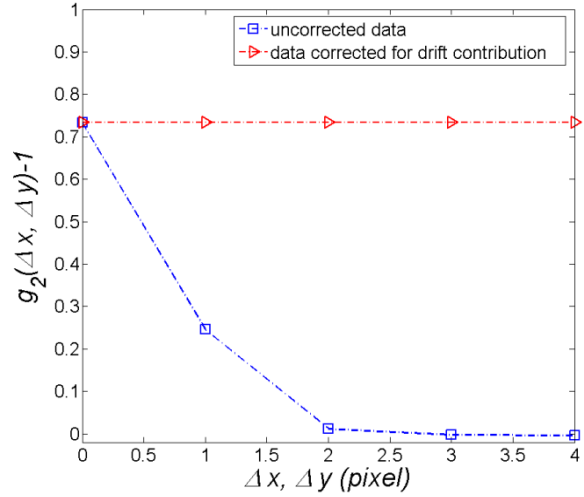

Figure 3. (Color online) Intensity correlation functions as a function of the imposed displacement for a time series of speckle images undergoing a rigid shift. The (numerically) applied displacement is 1pixel along $x$ and 1 pixel along $y$ per frame. The red triangles have been obtained using the drift-corrected formula, Eq (7), while the blue squares correspond to the usual formulas, Eqs. (5) and (6). The size of the ROI is 30×30 pixel² and $M = 2$.

An important point in the calculation of the corrected correlation function is the value of the factor $M$ which corresponds to the number of terms of the kernel function. We now show that this factor controls the accuracy of the correction scheme for subpixel displacements. We apply to the initial speckle image a series of consecutive displacements of 0.5 pixel along the $x$ direction and of 0.5 pixel along the $y$ direction, using a bilinear interpolation function available in Matlab (i.e. the function calculates the new pixel value as the weighted average of intensity of the four nearest pixels).

Figure 4 shows the corrected intensity correlation function as a function of the displacement for ROIs of size 30×30 pixel². The correlation function clearly oscillates: for half-integer displacements the correlation is lower than expected, while for integer displacements the correlation is identical to the auto-correlation of the initial image. As the factor $M$ is increased, the spurious loss of correlation due to sub-pixel shift is reduced. Note however that a large value of $M$ will lead to an unacceptable increase in computation time. As shown in Figure 4, $M$=8 to 10 appears to be a good compromise. In the following, we systematically use $M$ =10.

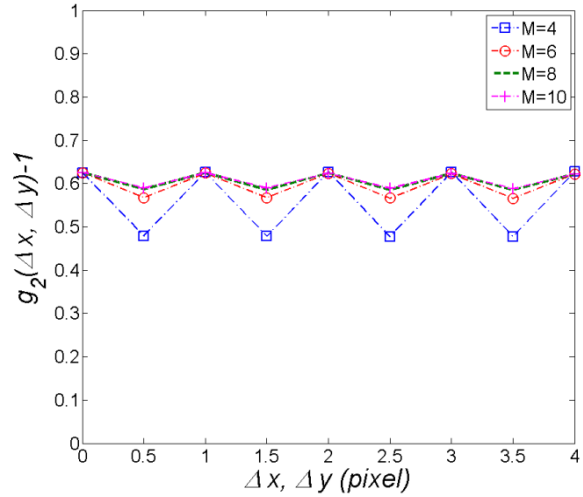

Figure 4. (Color online). Drift-corrected intensity correlation $g_2(\Delta x, \Delta y) - 1$ as a function of applied displacement, for different values of $M$. The applied displacement is 0.5 pixel along the $x$ direction and 0.5 pixel along the $y$ direction per frame. The size of ROI is 30×30 pixel$^2$ and speckle size is $S = 2.7$ pixels.

## 4 Experimental tests

In this section, we test our setup on two limiting cases: an elastic solid and a Newtonian liquid. For the former, we first check the consistency between an imposed macroscopic displacement and the displacement calculated from the speckle pattern. We then apply a constant shear strain, against which we compare the strain profile across the Couette gap calculated from the local drift of the speckle pattern. Finally, we apply a time-varying sinusoidal shear strain and check whether the measured sample response is consistent with linear elasticity. For the Newtonian liquid, we verify that the velocity profile extracted from the drift of the speckle pattern agrees with that inferred from the applied shear rate. Additionally, we investigate the (drift-corrected) intensity correlation functions probing the microscopic dynamics of the tracer particles, both at rest and under shear, confirming quantitatively the Newtonian character of the fluid.

### 4.1 Elastic solid

In order to evaluate the resolution, accuracy, and repeatability of our setup, we first evaluate its performance on an elastic solid. We use the reticulated PDMS seeded with 300 nm diameter Silica particles which produces a frozen speckle pattern. The particle's concentration is sufficiently small to neglect multiple scattering. By analyzing the speckle images using a Particle Image Velocimetry [26] [27] software (PIVlab) [28], we first measure the displacement field when the sample is uniformly displaced using a translational optical stage. We then measure the displacement field in response to a sinusoidal shear strain.

#### 4.1.1 Homogeneous displacement

Figure 5 (a) shows the micro-displacement map of different ROIs obtained for an applied displacement of 20 μm using a manual micrometric translation stage in place of the Couette cell geometry. The colors indicate the magnitude of displacement, ranging from blue (10 μm) to red (30 μm). As expected, the measured displacement is uniform throughout the scattering volume with an excellent homogeneity. The recorded average displacement is 19.87 μm, in excellent agreement with the applied displacement (20 μm). Figure 5(b) displays the Probability Density Function of the shifts for all ROIs. The continuous red line is a Gaussian fit with a variance of 0.13 μm², i.e. standard deviation of 0.36 μm, indicative of the accuracy of our setup in measuring a rigid displacement.

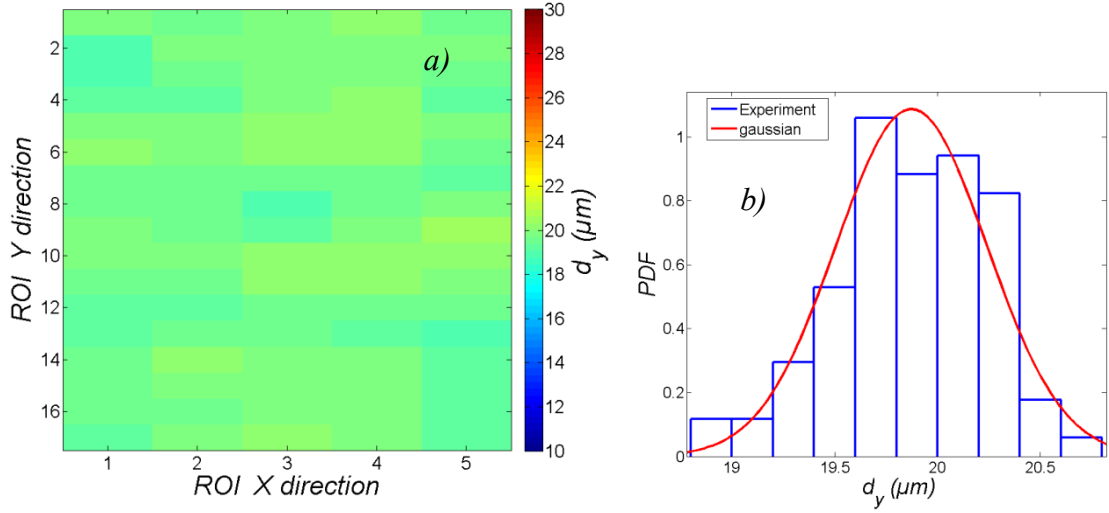

Figure 5. (Color online). a) micro-displacement map measured between two pairs of speckle images generated by a sample of PDMS. The first image is taken at rest and the second one after physically translating the sample by 20 μm. b) PDF (Probability Density Function) of the displacements in the y direction, obtained from the data for all ROIs.

*4.1.2  Step shear strain*

We now evaluate the performance of our setup in the case of a simple shear applied to an elastic solid. For that, a shear strain is applied to a PDMS sample filling the Couette geometry. We take a first speckle image of the gap at rest and then a second one after applying a shear strain $\gamma = 0.3\%$. The PIVlab software is used to measure the displacement field in the Couette cell gap.

Figure 6 shows the displacement $d_\theta(r)$ computed from the analysis of the speckle images. The red line is the theoretical shear strain profile:

$$d_\theta(r) = \frac{\theta R_i^2}{R_e^2 - R_i^2}(\frac{R_e^2}{r} - r),  \tag{8}$$

where $R_i$ is the radius of the inner cylinder, $R_e$ that of the outer one and $\theta$ is the deflection angle applied by the rheometer.

Figure 6 shows a good agreement between the experimental and theoretical profiles (Eq (8)). Moreover, the value of the displacement approaches zero in the vicinity of the stator, confirming that the PDMS sticks well to the wall and does not present any slip. The inset shows that the radial component of displacement is close to zero ($d_r(r) \approx 0$), as expected.

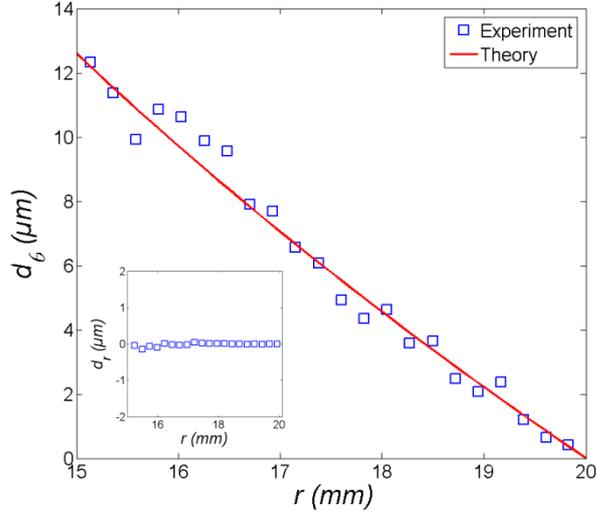

Figure 6. (Color online). Tangential displacement profile $d_\theta(r)$ measured between two pairs of speckle images generated by a sample of PDMS. The first image is taken at rest, the second one after applying a shear strain $\gamma = 0.3\%$. Inset: radial displacement profile $d_r(r)$ vs $r$.

### *4.1.3 Oscillatory shear strain*

The linear regime of an elastic material corresponds to the absence of evolution of its microscopic structure, besides the trivial affine deformation field. As a consequence, the speckle field should come back to its original pattern if a shear strain is applied and then removed. To verify that, a small oscillatory shear strain ($\gamma = 0.05$, $f = 0.4 Hz$) is applied, while recording speckle images at 20 frames/s during 90 seconds.

Figure 7 (a) shows the intensity autocorrelation function with no drift correction (Eq (6)), calculated for different ROIs from a time series of speckle images taken while the sample undergoes oscillatory shear strain. For all ROIs, $g_2(\tau) - 1$ exhibits oscillations that are perfectly synchronous with the imposed deformation, and recovers its initial value at the beginning of each period. This demonstrates that the sample recovers its microscopic structure, indicative of a fully linear material response at the microscopic scale. The amplitude of the oscillations of $g_2(\tau) - 1$ is maximal for the ROIs close to the rotor: this is related to the strain-induced relative motion of the scatterers that will be investigated in detail in subsection 4.2.2.2. For the ROIs close to the stator, the values of $g_2(\tau) - 1$ are almost constant, showing small oscillations that are the result of the scatterers average motion in the finite-sized ROI, whose width is 50 pixels.

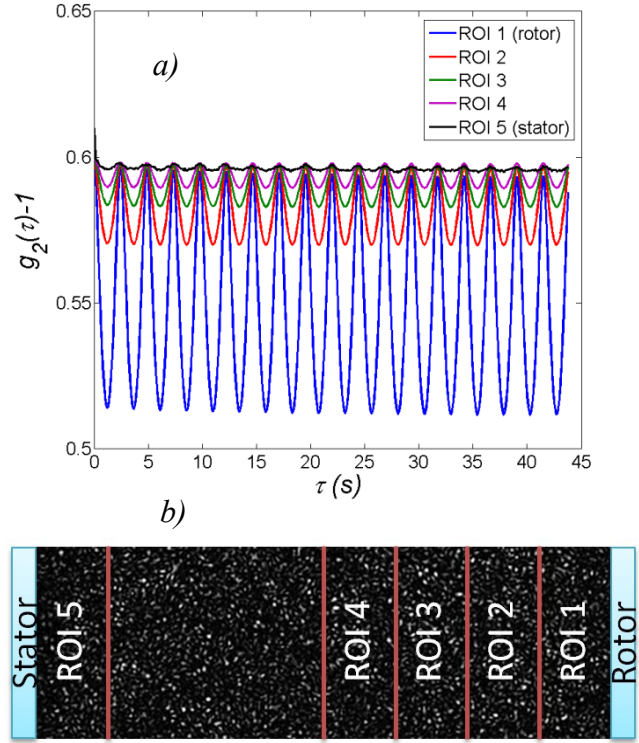

Figure 7. (Color online). a) Intensity correlation function $g_2(\tau)-1$ calculated from speckle images generated by an elastic solid (PDMS) under oscillatory strain ( $\gamma = 0.05\%$, $f = 0.4Hz$ ) in the linear domain. The different curves are measured at various locations in the gap, as shown in b). The ROI size is 50×150 pixel.

### 4.2 Newtonian fluid

To complete the experimental setup validation, we fill the Couette geometry with a Newtonian fluid seeded with small amounts of polystyrene Brownian particles (see subsection 2.1.2 for more details). We first determine the particle's diameter by measuring $g_2(\tau)-1$ at rest. We then shear the suspension by applying a constant shear rate while recording speckle images. For the sequence of images acquired under constant shear, we use the PIVlab software to calculate the velocity profile in the gap. We then calculate the drift-corrected intensity correlation function, using Eq (7).

#### 4.2.1 Microscopic dynamics at rest

Figure 8 shows the intensity correlation function, corrected for the dark noise and the non-uniform illumination as discussed in [29], for the Newtonian suspension at rest. The decay of $g_2(\tau)-1$ follows a single exponential, as expected for nearly monodisperse Brownian particles. The continuous red line is an exponential fit yielding a hydrodynamic diameter $D_h = 184nm$, which is in a very good agreement with $D_h = 180nm$ as quoted by the manufacturer. The quality of the fit can be better appreciated in the inset of the Figure 8 where $g_2(\tau)-1$ is plotted in a semilogarithmic graph: the correlation function follows a straight line over more than 3 decades. Deviations from this behavior at large $\tau$ are due to the instantaneous noise of the camera [29] whose contribution to $g_2(\tau)-1$ has a magnitude of about $10^{-4}$.

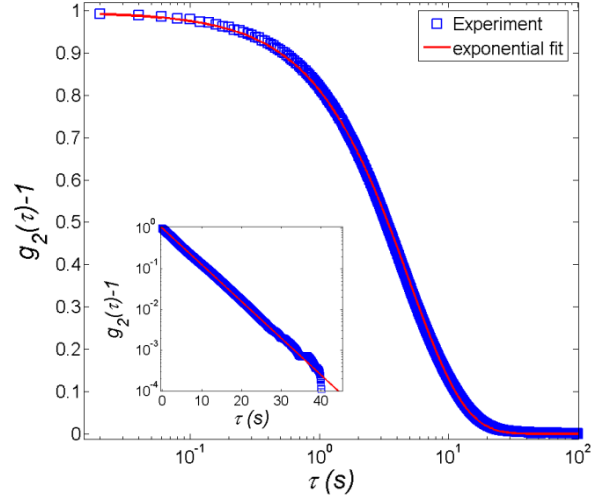

Figure 8. (Color online). Intensity correlation function as a function of time delay $\tau$ for Brownian particles (polystyrene nanospheres with hydrodynamic diameter $D_h = 180nm$) suspended in a Newtonian fluid (aqueous solution of Emkarox of viscosity $\eta = 11.4 Pa.s$ at 24°C). Data are averaged over $N = 10000$ pixels and over $T_{exp} = 200s$.

### 4.2.2 Steady shear flow

#### 4.2.2.1 Velocity profiles

Several constant shear rates were imposed to the fluid while recording speckle images with a frequency proportional to the applied shear rate. The inset of Figure 9 shows the velocity profiles in the gap of the Couette geometry for those shear rates. Velocity values obtained by PIV were averaged over 1s. The velocity in the gap is expected to be [30]:

$$v_\theta(r) = \frac{\Omega R_i^2}{R_e^2 - R_i^2}\left(\frac{R_e^2}{r} - r\right), \qquad (9)$$

where $\Omega$ is the angular velocity imposed by the rheometer.

Figure 9 is a non-dimensional representation of the velocity in the gap, showing a very good agreement between experimental data and theory (Eq (9)). Further, the velocity is close to zero in the vicinity of the outer cylinder ($r=20mm$) showing the absence of slip at the wall. Finally, a significant deviation from linearity is observed, showing that the shear rate is not constant in the gap, as indicated by Eq (9).

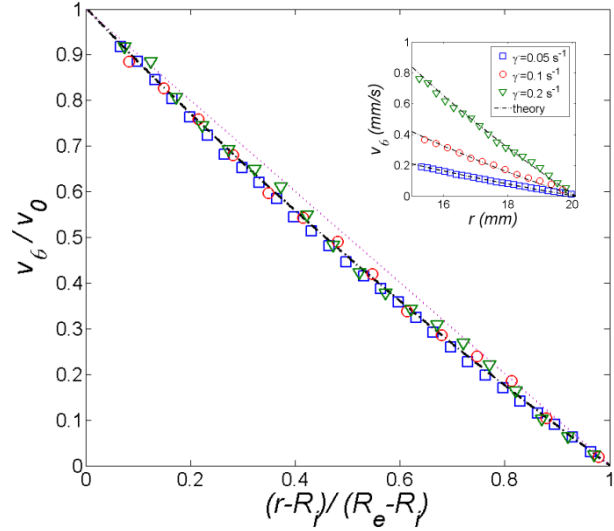

Figure 9. (Color online). Dimensionless shear velocity $v_\theta/v_0$ *vs* dimensionless radial coordinate $(r-R_i)/(R_e-R_i)$. Inset: Velocity profiles as a function of the radial coordinate in the gap of a Couette cell, for different shear rates. The velocity is averaged over a time of 1s. Same sample as in figure. 8. Black dot dashed line: theory (Eq (9)), purple dotted line: linear velocity profile for comparison.

*4.2.2.2 Microscopic dynamics under shear*

Since shear is a combination of solid rotation and stretching, we expect the intensity correlation function to contain contributions from both the rigid drift and the relative motion due to the deformation associated with stretching. For the sake of comparison, we thus compute both the standard $g_2(\tau)-1$ (Eq (6)) and the drift-correct correlation function (Eq (7)) using the velocity field extracted from the PIV analysis.

Figure 10 shows the intensity correlation function for the fluid at rest and sheared at $\dot{\gamma}=0.1 s^{-1}$, with and without the drift correction. Under shear, the characteristic decay time decreases by two decades for the uncorrected data and by one decade for the corrected ones. We emphasize that even after correction, $g_2(\tau)-1$ under shear does not superimpose on the curve for the sample at rest. This is due to the different origin of the microscopic dynamics (relative motion) in the two cases: at rest, the dynamics are due to Brownian motion, while under shear the relative displacement due to the deformation field adds up to the Brownian motion.

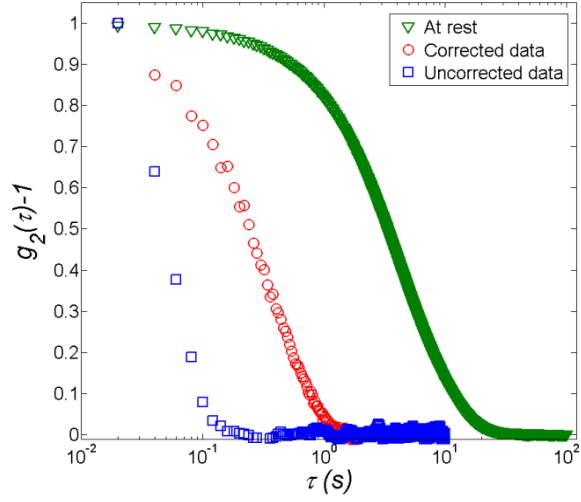

Figure 10. (Color online). Intensity correlation function $g_2(\tau)-1$ as a function of the time delay $\tau$ for the same suspension of Brownian particles in a Newtonian fluid as in figure. 9. $g_2(\tau)-1$ is measured at rest and under a constant shear rate $\dot\gamma = 0.1 s^{-1}$, with or without drift correction as indicated by the labels. The data are for a ROI located at *r=17.5mm*.

To understand the contribution of the deformation, we recall that in our imaging geometry each speckle corresponds to a small sample volume comprising many particles. When the sample is strained, the relative positions of the particles within this volume change proportionally to the strain value. Therefore, $g_2(\tau)-1$ decays significantly when the typical difference of displacement between two particles within this volume is such that $\Delta\vec{r}\cdot\vec{q} \approx 2\pi$ [20]. From this formula, we can calculate an estimate of the relaxation time $\tau_{shear}^{th} \approx 0.043\dot\gamma_l^{-1}$ (see Appendix for the calculation's details). Importantly, the relaxation time is expected to be inversely proportional to the applied shear rate.

As we can see in the inset of the Figure 11, for a given ROI $g_2(\tau)-1$ is indeed strongly dependent on the applied shear rate. Moreover, the correlation function decays significantly for $\tau_{shear}^{exp} \approx 0.04\dot\gamma_l^{-1}$, in excellent agreement with our estimate $\tau_{shear}^{th} \approx 0.043\dot\gamma_l^{-1}$. We note that the scaling of $g_2(\tau)-1$ when plotted vs $\dot\gamma\tau$ implies that, in the present case, the contribution of Brownian motion is negligible with respect to that due to the affine deformation. Indeed, Figure 10 shows that the decay time due to Brownian motion alone is of the order of 10 s, much larger than the longest decay time due to shear, which is on the order of 1 s (see inset of Figure 11).

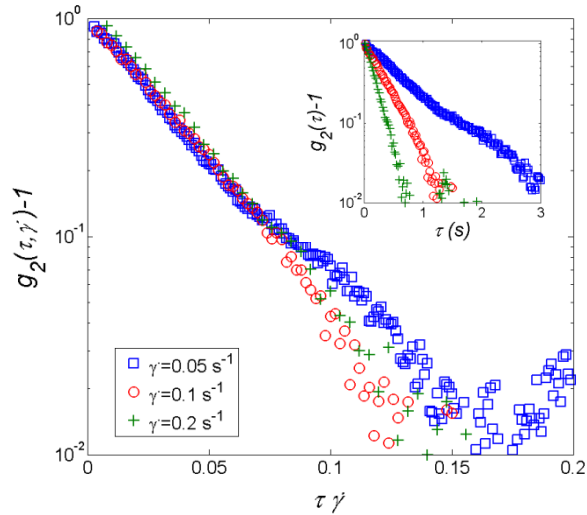

Figure 11. (Color online). Drift-corrected intensity correlation function $g_2(\tau)-1$ vs the normalized delay $\tau\dot\gamma$, for the same suspension of Brownian particles as in figures. 9 and 10. The correlation functions are measured for three values of the shear rate: $\dot\gamma = 0.05 s^{-1}$, $\dot\gamma = 0.1 s^{-1}$, and $\dot\gamma = 0.2 s^{-1}$, for a ROI of 32×32 pixel$^2$ located at *r=17.5mm*. Inset: Same data plotted as a function of the time delay $\tau$.

Finally, we check in Figure 12 that the above scaling is valid throughout the whole gap of the Couette cell, by measuring the drift-corrected $g_2(\tau)-1$ for 3 different ROIs of 32 x 32 pixel**2**, near the rotor (*r=15.7mm*), in the middle of the gap (*r=17.5mm*), and near the stator (*r=19.3 mm*), respectively. The inset of Figure 12 shows that when using the average applied shear rate, the rescaling is poor. Using the local shear rate measured by PIV, we find again a remarkable scaling onto a single master curve, confirming the analysis detailed above. This master curve can then be used to set an upper limit on the measurable characteristic time of additional material relaxations, that may add up to the relative motion due to affine deformation. An example of such dynamics are plastic rearrangements in amorphous solids, a topic of great current interest. Indeed, for a given shear rate $\dot\gamma$, any relaxation mechanism with a time scale (at the length scale $1/q$ probed by the setup) shorter than $\tau_{shear} \approx 4\pi\sqrt{2}/(qL\dot\gamma_l)$ (see Appendix) will measurably accelerate the decay of $g_2(\tau)-1$. The RheoSpeckle apparatus will therefore be particularly well suited to investigate the microscopic dynamics at very low shear e.g. during the slow creep of soft matter systems.

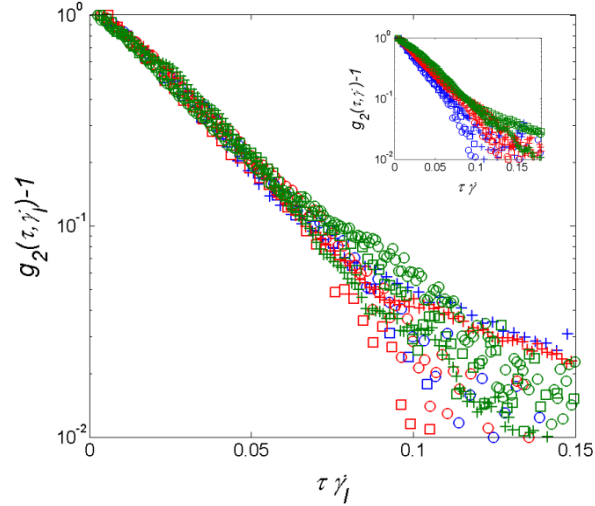

Figure 12. (Color online). Intensity correlation function $g_2(\tau)-1$ as a function of the delay $\tau$ normalized by the value of local shear rate $\dot\gamma_l$ for a suspension of Brownian particles. The $g_2(\tau)-1$ functions are measured for three values of the shear rate $\dot\gamma = 0.05 s^{-1}$ (squares), $\dot\gamma = 0.1 s^{-1}$ (circles) and $\dot\gamma = 0.2 s^{-1}$ (pluses), in 3 different ROIs of the gap: blue symbols, near the rotor (*r=15.7mm*); red symbols, in the middle of the gap (*r=17.5mm*); green symbols, near the stator (*r=19.3mm*). Inset: same data, with $\tau$ normalized by the average shear rate $\dot\gamma$.

## 5    Conclusions

We have presented a new setup that couples conventional rheometry to space and time-resolved light scattering. The novel features of our setup are its ability to measure simultaneously the macroscopic rheological behavior, the local velocity, and the local microscopic dynamics of the material. The setup has been successfully tested using two ideal materials: an elastic solid and a Newtonian fluid. For the elastic solid sample, rigid translations are measured with an accuracy better than 1 μm. Moreover, the rheological linear regime is found to also apply at the microscopic level, in that the initial microscopic configuration of the sample is perfectly recovered when the applied shear stress is removed.

The local flow of a Newtonian fluid has been measured over the full width of the gap (5 mm), with a temporal resolution of 1s and a spatial resolution, given by the ROI size, of 100 μm. When the fluid is at rest, the microscopic dynamics of tracer particles is found to be quantitatively consistent with Brownian diffusion. Under shear, the decay of the correlation function is dominated by the contribution due to stretching; accordingly, the decay rate of the correlation function is found to be proportional to the applied shear rate.

Our setup opens the field to a better understanding of the behavior of transparent complex fluids, in particular for glassy or jammed soft materials where slow dynamics are coupled to the effects of an external stress.

**Acknowledgments**

The Laboratoire Rhéologie et Procédés is part of the LabEx Tec 21 (Investissements d'Avenir - grant agreement n°ANR-11-LABX-0030) and of the PolyNat Carnot Institut (Investissements d'Avenir - grant agreement n°ANR-11-CARN-030-01). LC acknowledges funding from the ANR (project FAPRES, ANR-14-CE32-0005-01).

**Appendix: decay rate resulting from an affine shear deformation**

We evaluate the term $\Delta \vec{r}.\vec{q}$ discussed in subsection 4.2.2.2 for a shear deformation at constant shear rate. We start by calculating $\Delta \vec{r}$, the typical difference of displacement between two particles along the direction of $\vec{q}$. To estimate $\Delta \vec{r}$, we rely on the fact that the rate of deformation $\Gamma$ can be decomposed into a symmetric tensor $E$ (stretch) and a skew-symmetric tensor $\Omega$ (solid rotation) as follows [30]:

$$\Gamma = 1/2\left(\nabla \vec{v} + \nabla \vec{v}^T\right) + 1/2\left(\nabla \vec{v} - \nabla \vec{v}^T\right) = E + \Omega \tag{A1}$$

$$\Gamma = \begin{pmatrix} 0 & \dot{\gamma}_l & 0 \\ 0 & 0 & 0 \\ 0 & 0 & 0 \end{pmatrix} = \begin{pmatrix} 0 & \dot{\gamma}_l/2 & 0 \\ \dot{\gamma}_l/2 & 0 & 0 \\ 0 & 0 & 0 \end{pmatrix} + \begin{pmatrix} 0 & \dot{\gamma}_l/2 & 0 \\ -\dot{\gamma}_l/2 & 0 & 0 \\ 0 & 0 & 0 \end{pmatrix} \tag{A2}$$

We recall that $g_2(\tau) - 1$ shown in figures 11 and 12 is corrected from the contribution of solid rotation, so that the decay of $g_2(\tau) - 1$ is the result of the relative displacement due to the stretch tensor $E$ only. Now, let's consider two particles belonging to a scattering volume $V$ of typical size $L$ that contributes to a given speckle. When applying a shear at a rate $\dot{\gamma}$ during a time $\tau_{shear}^{th}$, the scattering volume is deformed by $\tau_{shear}^{th} \underline{\underline{E}} \cdot \vec{L}$, where $\vec{L}$ is the characteristic vector of $V$. Since the displacement field has no $z$ component, the relative displacement $\Delta \vec{r}$ is:

$$\Delta \vec{r} = \begin{pmatrix} \Delta r_x \\ \Delta r_y \\ \Delta r_z \end{pmatrix} = \tau_{shear}^{th} \begin{pmatrix} 0 & \dot{\gamma}_l/2 & 0 \\ \dot{\gamma}_l/2 & 0 & 0 \\ 0 & 0 & 0 \end{pmatrix} \begin{pmatrix} L \\ L \\ L \end{pmatrix} \tag{A3}$$

$$\Delta \vec{r} = \begin{pmatrix} \Delta r_x \\ \Delta r_y \\ \Delta r_z \end{pmatrix} = \begin{pmatrix} (\dot{\gamma}_l/2)L\tau_{shear}^{th} \\ (\dot{\gamma}_l/2)L\tau_{shear}^{th} \\ 0 \end{pmatrix} \tag{A4}$$

As the experimental scattering angle is $\theta = 90°$, the scattering vector has components only along the directions of the incident ($x$) and scattered ($z$) light, with $q_x = q_z = q/\sqrt{2}$. It follows that:

$$\Delta \vec{r} \cdot \vec{q} = \begin{pmatrix} (\dot{\gamma}_l/2)L\tau_{shear}^{th} \\ (\dot{\gamma}_l/2)L\tau_{shear}^{th} \\ 0 \end{pmatrix} \begin{pmatrix} q_x \\ 0 \\ q_z \end{pmatrix} = (\dot{\gamma}_l/2\sqrt{2})L\tau_{shear}^{th} q \quad (A5)$$

Finally, the correlation function relaxes significantly for $\Delta \vec{r} \cdot \vec{q} = \Delta r_x \cdot q_x \approx 2\pi$. From Eq (A5), one has:

$$\dot{\gamma}_l \tau_{shear}^{th} \approx 4\pi\sqrt{2}/(qL) \quad (A6)$$

Using, $q = 22 \mu m^{-1}$ and $L=$ *speckle size$=19.04$ μm*, the speckle size in our imaging geometry, we find $\dot{\gamma}_l \tau_{shear}^{th} \approx 0.043$, the result quoted in subsection 4.2.2.2.